\documentclass[nofootinbib,twocolumn,showpacs,preprintnumbers,pre,aps]{revtex4-1}

\usepackage{graphicx}
\usepackage{bm}
\usepackage{amsmath}
\usepackage{amssymb}
\usepackage{color}

\begin{document}

\title{Elastic three-sphere microswimmer in a viscous fluid}%

\author{Kento Yasuda}

\author{Yuto Hosaka}

\author{Mizuki Kuroda}

\author{Ryuichi Okamoto}

\author{Shigeyuki Komura}\email{komura@tmu.ac.jp}

\affiliation{
Department of Chemistry, Graduate School of Science and Engineering,
Tokyo Metropolitan University, Tokyo 192-0397, Japan}

\date{\today}

\begin{abstract}
We discuss the dynamics of a generalized three-sphere microswimmer in which the spheres 
are connected by two elastic springs.
The natural length of each spring is assumed to undergo a prescribed cyclic change.
We analytically obtain the average swimming velocity as a function of the frequency of 
the cyclic change in the natural length.
In the low-frequency region, the swimming velocity increases with the frequency and its 
expression reduces to that of the original three-sphere model by Najafi and Golestanian.
In the high-frequency region, conversely, the average velocity decreases with 
increasing the frequency. 
Such a behavior originates from the intrinsic spring relaxation dynamics of an elastic swimmer 
moving in a viscous fluid. 
\end{abstract}

\maketitle

Microswimmers are tiny machines that swim in a fluid such as sperm cells or motile bacteria, 
and are expected to be applied to microfluidics and microsystems~\cite{Lauga092}.
By transforming chemical energy into mechanical work, microswimmers change their 
shape and move in viscous environments.
Over the length scale of microswimmers, the fluid forces acting on them are governed by the 
effect of viscous dissipation.
According to Purcell's scallop theorem~\cite{Purcell77}, time-reversal body motion 
cannot be used for locomotion in a Newtonian fluid~\cite{Lauga11}.
As one of the simplest models exhibiting broken time-reversal motion, Najafi and Golestanian 
proposed a three-sphere swimmer~\cite{Golestanian04,Golestanian08}, where three 
in-line spheres are linked by two arms of varying length.
This model is suitable for analytical analysis because it is sufficient to consider only the 
translational motion and the tensorial structure of the fluid motion can be neglected.
Recently, such a swimmer has been experimentally realized by using ferromagnetic particles 
at an air-water interface and applying an oscillating magnetic field~\cite{Grosjean}.

The original Najafi--Golestanian model has been further extended to various different 
cases such as when one of the spheres has a larger radius~\cite{Golestanian2008}, or 
when three spheres are arranged in a triangular configuration~\cite{Aguilar12}.
Montino and DeSimone considered the case in which one arm is periodically actuated 
while the other is replaced by a passive elastic spring~\cite{Montino15}.
It was shown that such a swimmer exhibits a delayed mechanical response of the 
passive spring with respect to the active arm.
More recently, they analyzed the motion of a three-sphere swimmer whose arms 
have active viscoelastic properties mimicking muscular contraction~\cite{Montino17}.

Another way of extending the Najafi--Golestanian model is to consider the arm motions 
to occur stochastically~\cite{GolestanianAjdari08,Golestanian10}, rather than 
assuming a prescribed sequence of deformations~\cite{Golestanian04,Golestanian08}.
In these models, the configuration space of a swimmer generally consists of finite number 
of distinct states. 
A similar idea was employed by Sakaue \textit{et al.}\ who discussed  propulsion of 
molecular machines or active proteins in the presence of hydrodynamic interactions~\cite{Sakaue10}. 
Later Huang \textit{et al.}\ considered a modified three-sphere swimmer in a two-dimensional 
viscous fluid~\cite{Huang12}.
In their model, the spheres are connected by two springs whose lengths are assumed to 
depend on the discrete states that are cyclically switched. 
As a result, the dynamics of a swimmer consists of the spring relaxation processes which 
follow after each switching event.

\begin{figure}[b]
\begin{center}
\includegraphics[scale=0.35]{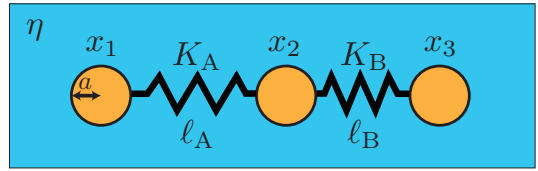}
\end{center}
\caption{
Elastic three-sphere microswimmer in a viscous fluid characterized by the shear viscosity 
$\eta$. 
Three identical spheres of radius $a$ are connected by two harmonic springs whose 
elastic constants are $K_{\rm A}$ and $K_{\rm B}$.  
The natural lengths of the springs, $\ell_{\rm A}(t)$ and $\ell_{\rm B}(t)$, depend on 
time and are assumed to undergo cyclic change [see Eqs.~(\ref{ellA}) and (\ref{ellB})].
The time-dependent positions of the spheres are denoted by $x_1(t)$, $x_2(t)$, and 
$x_3(t)$ in a one-dimensional coordinate. 
}
\label{model}
\end{figure}

In this letter, we discuss a generalized three-sphere swimmer in which the spheres 
are simply connected by two harmonic springs.  
The main difference compared with the previous models is that the natural length 
of each spring depends on time and is assumed to undergo a prescribed cyclic change. 
Whereas the arms in the Najafi--Golestanian model undergo a prescribed motion 
regardless of the force exerted by the fluid, the sphere motion in our model is 
determined by the natural spring lengths representing internal states of a swimmer, 
and also by the force exerted by the fluid. 
In this sense, our model is more realistic to study the locomotion of active microswimmers. 
We analytically obtain the average swimming velocity as a function of the frequency of 
the cyclic change in the natural length.
In order to better illustrate our result, we first explain the case when the two spring 
constants are identical, and also the two oscillation amplitudes of the natural lengths are 
the same.
Then we shall argue a general case when these quantities are different and when the 
phase mismatch between the natural lengths is arbitrary.

The introduction of harmonic springs between the spheres leads to an intrinsic time scale 
of an elastic swimmer that  characterizes its internal relaxation dynamics. 
When the frequency of the cyclic change in the natural lengths is smaller than this 
characteristic time, the swimming velocity increases with the frequency as 
in the previous works~\cite{Golestanian08}.
In the high-frequency region, on the other hand, the motion of spheres cannot follow 
the change in the natural length, and the average swimming velocity decreases with 
increasing the frequency. 
Such a situation resembles to the dynamics of the Najafi--Golestanian three-sphere swimmer 
in a viscoelastic medium~\cite{Yasuda17}. 
We also show that, due to the elasticity that has been introduced, the proposed micromachine 
can swim even if the change in the natural lengths is reciprocal as long as its structural 
symmetry is violated.
Although the considered swimmer appears to be somewhat trivial, it can be regarded as a 
generic model for microswimmers or protein machines since the behaviors of the previous 
models can be deduced from our model by taking different limits.

We generalize the Najafi--Golestanian three-sphere swimmer model to take into account 
the elasticity in the sphere motion.
As schematically shown in Fig.~\ref{model}, the present model consists of three hard spheres 
of the same radius $a$ connected by two harmonic springs A and B whose spring constants 
are $K_{\rm A}$ and $K_{\rm B}$, respectively.
We assume that the natural lengths of these springs, denoted by $\ell_{\rm A}(t)$ and 
$\ell_{\rm B}(t)$, undergo cyclic time-dependent change.
Their explicit time dependences will be specified later.
The total energy of an elastic swimmer is then given by 
\begin{align}
E = \frac{K_{\rm A}}{2}(x_2 - x_1 - \ell_{\rm A})^2 + 
\frac{K_{\rm B}}{2}(x_3 - x_2 - \ell_{\rm B})^2,
\end{align}
where $x_i(t)$ ($i=1, 2, 3$) are the positions of the three spheres in a one-dimensional 
coordinate. 
We also assume $x_1<x_2<x_3$ without loss of generality.
Owing to the hydrodynamic interaction, each sphere exerts a force on the viscous fluid 
of shear viscosity $\eta$ and experiences an opposite force from it.
In general, the surrounding medium can be viscoelastic~\cite{Yasuda17}, but such an 
effect is not considered in this letter.

Denoting the velocity of each sphere by $\dot x_i$, we can write the equations of motion 
of the three spheres  as 
\begin{widetext}
\begin{align}
\dot x_1&=\frac{K_{\rm A}}{6\pi\eta a}(x_2-x_1-\ell_{\rm A})
-\frac{K_{\rm A}}{4\pi\eta}\frac{(x_2-x_1-\ell_{\rm A})}{x_2-x_1}
+\frac{K_{\rm B}}{4\pi\eta}\frac{(x_3-x_2-\ell_{\rm B})}{x_2-x_1}
-\frac{K_{\rm B}}{4\pi\eta}\frac{(x_3-x_2-\ell_{\rm B})}{x_3-x_1},  
\label{V1} \\
\dot x_2 &=\frac{K_{\rm A}}{4\pi\eta}\frac{(x_2-x_1-\ell_{\rm A})}{x_2-x_1}
-\frac{K_{\rm A}}{6\pi\eta a}(x_2-x_1-\ell_{\rm A})+
\frac{K_{\rm B}}{6\pi\eta a}(x_3-x_2-\ell_{\rm B})
-\frac{K_{\rm B}}{4\pi\eta}\frac{(x_3-x_2-\ell_{\rm B})}{x_3-x_2},  
\label{V2} \\
\dot x_3 &=\frac{K_{\rm A}}{4\pi\eta}\frac{(x_2-x_1-\ell_{\rm A})}{x_3-x_1}
-\frac{K_{\rm A}}{4\pi\eta}\frac{(x_2-x_1-\ell_{\rm A})}{x_3-x_2}
+\frac{K_{\rm B}}{4\pi\eta}\frac{(x_3-x_2-\ell_{\rm B})}{x_3-x_2}
-\frac{K_{\rm B}}{6\pi\eta a}(x_3-x_2-\ell_{\rm B}),
\label{V3} 
\end{align}
\end{widetext}
where we have used the Stokes' law for a sphere and the Oseen tensor in a three-dimensional 
viscous fluid.
The swimming velocity of the whole object can be obtained by averaging the velocities of 
the three spheres:
\begin{align}
V = \frac{1}{3}(\dot x_1+\dot x_2+\dot x_3).
\end{align}
One of the advantages of the present formulation is that the motion of the spheres is simply 
described by coupled ordinary differential equations. 
Moreover, the force-free condition for the whole system~\cite{Golestanian04,Golestanian08} 
is automatically satisfied in the above equations.

Next we assume that the two natural lengths of the springs undergo the following periodic changes: 
\begin{align}
\ell_{\rm A}(t) & =\ell+d_{\rm A}\cos(\Omega t),  
\label{ellA} \\
\ell_{\rm B}(t) & =\ell+d_{\rm B}\cos(\Omega t - \phi).
\label{ellB}
\end{align}
In the above, $\ell$ is the common constant length, $d_{\rm A}$ and $d_{\rm B}$ are the amplitudes of 
the oscillatory change, $\Omega$ is the common frequency, and $\phi$ is the mismatch in the phases 
between the two cyclic changes.
The time-reversal symmetry of the spring dynamics is present only when $\phi=0$ or $\pi$, 
otherwise the time-reversal symmetry is broken.
In the following analysis, we generally assume that $d_{\rm A}, d_{\rm B}, a \ll \ell$ and focus 
on the leading order contribution. 
It is convenient to introduce a characteristic time scale $\tau=6\pi\eta a/K_{\rm A}$. 
Then we use $\ell$ to scale all the relevant lengths ($x_i$, $a$, $d_{\rm A}$, $d_{\rm B}$),
and employ $\tau$  to scale the frequency, i.e., $\hat \Omega = \Omega \tau$.
By further defining the ratio between the two spring constants as  
$\lambda = K_{\rm B}/K_{\rm A}$, the coupled Eqs.~(\ref{V1})--(\ref{V3}) can be made 
dimensionless.

In order to present the essential outcome of the present model, we shall first consider the 
simplest symmetric case, i.e., $\lambda=1$, 
$d_{\rm A}=d_{\rm B}=d$, and $\phi = \pi/2$.
Hence Eq.~(\ref{ellB}) now reads $\ell_{\rm B}(t)  =\ell+d\sin (\Omega t)$. 
For our later calculation, it is useful to introduce the following spring lengths with respect to $\ell$:
\begin{align}
u_{\rm A} =x_2-x_1-\ell,~~~~~
u_{\rm B} =x_3-x_2-\ell.
\label{uAuB}
\end{align}
Notice that these quantities are related to the sphere velocities in Eqs.~(\ref{V1})--(\ref{V3}) as 
\begin{align}
\dot u_{\rm A} = \dot x_2- \dot x_1,~~~~~ 
\dot u_{\rm B} = \dot x_3- \dot x_2. 
\label{dotu1dotu2}
\end{align}
Using Eqs.~(\ref{V1})--(\ref{V3}) and solving Eq.~(\ref{dotu1dotu2}) in the frequency 
domain, we obtain after the inverse Fourier transform as  
\begin{align}
u_{\rm A}(t)& \approx  \frac{9-3\hat\Omega+5\hat \Omega^2+ \hat \Omega^3}
{9+10 \hat \Omega^2+ \hat \Omega^4} d \cos(\Omega t) \nonumber \\
& +\frac{6 \hat \Omega-4 \hat\Omega^2+2 \hat \Omega^3}
{9+10 \hat \Omega^2+ \hat \Omega^4} d \sin(\Omega t),
\label{u1}
\end{align}
\begin{align}
u_{\rm B}(t) & \approx -\frac{6 \hat \Omega+4 \hat \Omega^2+2 \hat \Omega^3}
{9+10 \hat \Omega^2+ \hat \Omega^4} d \cos(\Omega t) \nonumber \\
& +\frac{9+3\hat \Omega +5 \hat \Omega^2- \hat \Omega^3}
{9+10 \hat \Omega^2+ \hat \Omega^4} d \sin(\Omega t),
\label{u2}
\end{align}
where we have used $a/\ell \ll 1$.

According to the calculation by Golestanian and Ajdari~\cite{Golestanian08}, the average 
swimming velocity of a three-sphere swimmer can generally be expressed up to the 
leading order in $u_{\rm A}/\ell$ and $u_{\rm B}/\ell$ as 
\begin{align}
\overline V = \frac{7a}{24 \ell^2} 
\langle u_{\rm A} \dot u_{\rm B} - \dot u_{\rm A} u_{\rm B} \rangle,
\label{GolbarV}
\end{align}
where the averaging $\langle \cdots \rangle$ is performed by time integration in a full cycle.
The above expression indicates that the average velocity is determined by the area enclosed 
by the orbit of the periodic motion in the configuration space~\cite{Golestanian08}.
Using Eqs.~(\ref{u1}) and (\ref{u2}) for an elastic microswimmer with $d/\ell, a/\ell \ll 1$, 
we obtain the lowest order contribution as 
\begin{align}
\overline V= \frac{7d^2a}{24\ell^2 \tau} 
\frac{3\hat \Omega(3+ \hat \Omega^2)}{9+10 \hat \Omega^2+ \hat \Omega^4},
\label{symvelocity}
\end{align}
which is an important result of this letter.

\begin{figure}[tbh]
\begin{center}
\includegraphics[scale=0.35]{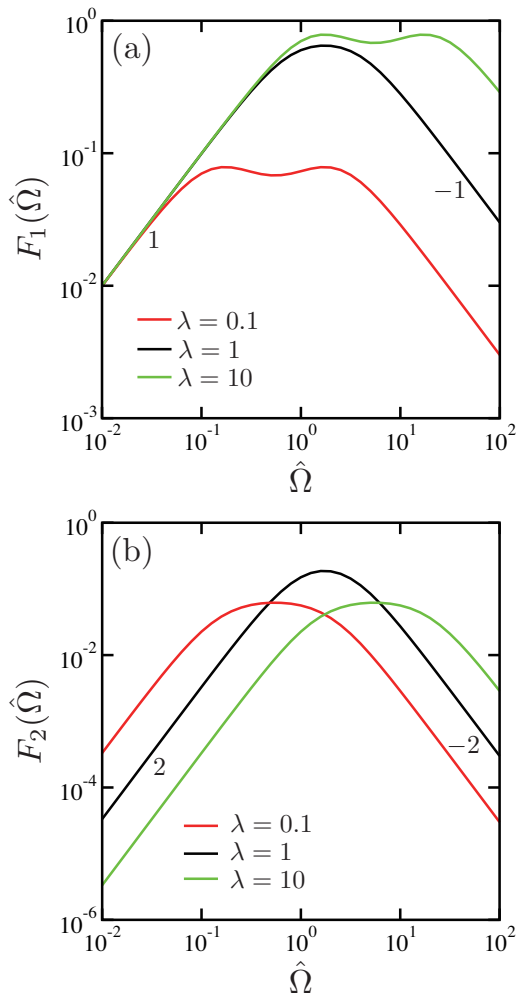}
\end{center}
\caption{
Plots of the scaling functions (a) $F_1(\hat \Omega; \lambda)$ and (b) 
$F_2(\hat \Omega; \lambda)$ defined in Eqs.~(\ref{F1}) and (\ref{F2}), respectively, 
as a function of $\hat \Omega = \Omega \tau$ for $\lambda= K_{\rm B}/K_{\rm A}=0.1, 1,$ and $10$.
The numbers indicate the slope representing the exponent of the power-law behaviors.
}
\label{scaling}
\end{figure}

We first consider the small-frequency limit of $\hat \Omega \ll 1$. 
Physically, this limit corresponds to the case when the spring constant $K_{\rm A}$ is 
very large. 
We easily obtain 
\begin{align}
u_{\rm A}(t) \approx d \cos(\Omega t),~~~~~
u_{\rm B}(t) \approx d \sin(\Omega t),
\label{u12small}
\end{align}
and 
\begin{align}
\overline V \approx \frac{7d^2a\Omega}{24\ell^2},
\label{barVsmall}
\end{align}
which exactly coincides with the average velocity of the Najafi-Golestanian 
swimmer with equal spheres~\cite{Golestanian04,Golestanian08}.
This is reasonable because the two spring lengths $u_{\rm A}$ and $u_{\rm B}$ 
are in phase with their respective natural lengths $\ell_{\rm A}$ and $\ell_{\rm B}$, 
as we see from Eqs.~(\ref{ellA}), (\ref{ellB}), and (\ref{u12small}).
Notice that the average velocity increases as $\overline V \sim \Omega$ in this limit, 
while it does not depend on the fluid viscosity $\eta$~\cite{Golestanian04,Golestanian08}.

In the opposite large-frequency limit of $\hat \Omega \gg 1$, on the other hand, we have   
\begin{align}
u_{\rm A}(t) & \approx \frac{\sqrt{5}d}{\Omega\tau} \cos[\Omega t-\arctan 2],  \\
u_{\rm B}(t) & \approx \frac{\sqrt{5}d}{\Omega\tau} \sin[\Omega t-(\pi-\arctan2)], 
\end{align}
where $\arctan 2 \approx 1.107$ and 
\begin{align}
\overline V \approx\frac{21d^2a}{24\ell^2 \Omega \tau^2}.
\label{barVlarge}
\end{align}
We see here that $u_{\rm A}$ and $u_{\rm B}$ are out of phase with respect to 
the natural lengths $\ell_{\rm A}$ and $\ell_{\rm B}$, while the average velocity 
decreases as $\overline V \sim \Omega^{-1}$ when $\Omega$ is increased.
When the spring constant $K_{\rm A}$ is small, it takes time for a spring to relax to its 
natural length, which leads to a delay in the mechanical response. 
The crossover frequency between the above two regimes is determined by $\hat \Omega \sim 1$. 
The general frequency dependence of Eq.~(\ref{symvelocity}) is shown in Fig.~\ref{scaling}(a) 
for $\lambda=1$ (black line). 
It shows a maximum around $\hat \Omega \sim 1$, as expected.

Recently, we have investigated the motion of the Najafi-Golestanian three-sphere 
swimmer in a viscoelastic medium~\cite{Yasuda17}.
We derived a relation that connects the average swimming velocity and the 
frequency-dependent viscosity of the surrounding medium. 
In this relation, the viscous contribution can exist only when the time-reversal symmetry 
is broken, whereas the elastic contribution is present only when the structural symmetry 
of the swimmer is broken.
In particular, we calculated the average swimming velocity when the surrounding 
viscoelastic medium is described by a simple Maxwell fluid with a characteristic time 
scale $\tau_{\rm M}$.
It was show that the viscous term increases as $\overline{V} \sim \Omega$ for 
$\Omega \tau_{\rm M} \ll 1$, while it decreases as $\overline{V} \sim \Omega^{-1}$ 
for $\Omega \tau_{\rm M} \gg 1$.
This is a unique feature of a swimmer in a viscoelastic medium~\cite{Yasuda17,Lauga09,Curtis13}, 
and such a reduction occurs simply because the medium responds elastically in the high-frequency 
regime. 
We note that the frequency dependence of $\overline V$ for an elastic three-sphere swimmer, 
as obtained in Eqs.~(\ref{symvelocity}), is analogous to the Najafi-Golestanian swimmer in 
a Maxwell fluid. 
In other words, an elastic microswimmer in a viscous fluid exhibits ``viscoelastic" effects as a whole.

Having discussed the simplest situation of the proposed elastic swimmer,  
we now show the result for a general case when $K_{\rm A} \neq K_{\rm B}$ 
(or $\lambda \neq 1$),  
$d_{\rm A} \neq d_{\rm B}$ and the phase mismatch $\phi$ in Eq.~(\ref{ellB}) is arbitrary. 
By repeating the same calculation as before, the spring lengths in Eq.~(\ref{uAuB}) now become 
\begin{align}
u_{\rm A}(t) \approx & \frac{1}{9\lambda^2+2(2+\lambda+2\lambda^2)\hat\Omega^2+\hat\Omega^4} 
\nonumber \\
\times & \biggl\{ [9\lambda^2+(4+\lambda)\hat\Omega^2]d_{\rm A} \cos(\Omega t) \nonumber  \\
& +2(3\lambda^2+\hat\Omega^2)\hat\Omega d_{\rm A} \sin(\Omega t) \nonumber \\
& - 2\lambda(1+\lambda)\hat\Omega^2 d_{\rm B} \cos(\Omega t-\phi) \nonumber \\ 
& - \lambda(-3\lambda+\hat\Omega^2)\hat\Omega d_{\rm B} \sin(\Omega t-\phi) \biggr\}, 
\end{align}
\begin{align}
u_{\rm B}(t) \approx & \frac{1}{9\lambda^2+2(2+\lambda+2\lambda^2)\hat\Omega^2+\hat\Omega^4} 
\nonumber \\
\times & \biggl\{-2 (1+\lambda)\hat\Omega^2 d_{\rm A} \cos(\Omega t) \nonumber \\
& + (3\lambda-\hat\Omega^2)\hat\Omega d_{\rm A} \sin(\Omega t) \nonumber \\
& + \lambda[9\lambda+(1+4\lambda)\hat\Omega^2] d_{\rm B} \cos(\Omega t-\phi) \nonumber \\
& + 2 \lambda(3+\hat\Omega^2)\hat\Omega d_{\rm B} \sin(\Omega t-\phi) \biggr\},
\end{align}
respectively, where we have used $a/\ell \ll 1$.
Using again Eq.~(\ref{GolbarV}), we finally obtain the lowest order general expression of the average 
velocity as  
\begin{align}
\overline V & =\frac{7 d_{\rm A}d_{\rm B} a}{24\ell^2\tau}
F_1(\hat \Omega;\lambda)\sin\phi
\nonumber \\
& -\frac{7 (\lambda-1)d_{\rm A} d_{\rm B} a}{12\ell^2 \tau}
F_2(\hat \Omega;\lambda) \cos\phi
\nonumber \\
& +\frac{7(d_{\rm A}^2-d_{\rm B}^2\lambda) a}{24\ell^2\tau}
F_2(\hat \Omega;\lambda),
\label{generalvelocity}
\end{align}
where the two scaling functions are defined by 
\begin{align}
F_1(\hat \Omega;\lambda) &=
\frac{3\lambda \hat \Omega(3\lambda+\hat \Omega^2)}
{9\lambda^2+2(2+\lambda+2\lambda^2)\hat \Omega^2+\hat \Omega^4},
\label{F1}
\\
F_2(\hat \Omega;\lambda) & 
=\frac{3\lambda \hat\Omega^2}{9\lambda^2+2(2+\lambda+2\lambda^2)\hat \Omega^2+\hat \Omega^4}.
\label{F2}
\end{align}
In Fig.~\ref{scaling}, we plot the above scaling functions as a function of $\hat \Omega$ for 
different $\lambda$ values.

When $\lambda=1$, $d_{\rm A}=d_{\rm B}$, and $\phi = \pi/2$, only the first term 
remains, and Eq.~(\ref{generalvelocity}) reduces to Eq.~(\ref{symvelocity}) as it should. 
When $\lambda \neq 1$, on the other hand, the second term is present even if $\phi=0$.
The third term is also present when $d_{\rm A}^2 \neq d_{\rm B}^2\lambda$ regardless 
of the phase mismatch $\phi$.
Notice that both the second and the third terms reflect the structural asymmetry of an 
elastic three-sphere swimmer, whereas the first term represents the broken time-reversal 
symmetry for $\phi \neq 0$~\cite{Yasuda17}. 
It is interesting to note that the frequency dependence of the second and the third terms
in Eq.~(\ref{generalvelocity}), represented by $F_2(\hat \Omega, \lambda)$, is different 
from that of the first term, represented by $F_1(\hat \Omega, \lambda)$.
According to Eq.~(\ref{F2}), $\overline V$ due to the second and the third terms increases 
as $\overline{V} \sim \Omega^2$ for $\hat \Omega \ll 1$, whereas it decreases as 
$\overline{V} \sim \Omega^{-2}$ for $\hat \Omega \gg 1$.
In general, the overall swimming velocity depends on various structural parameters and 
exhibits a complex frequency dependence. 
For example, we point out that $F_1(\hat \Omega, \lambda)$ in Fig.~\ref{scaling}(a) 
exhibits non-monotonic frequency dependence (two maxima) for $\lambda=0.1$ or $10$ 
(namely, when $\lambda \neq 1$). 
On the other hand, an important common feature in all the terms in Eq.~(\ref{generalvelocity}) 
is that $\overline V$ decreases for $\hat \Omega \ge 1$, which is characteristic for 
elastic swimmers.

We confirm again that Eq.~(\ref{generalvelocity}) reduces to the result by Golestanian and 
Ajdari~\cite{{Golestanian08}}, i.e., $\overline V =7 d_{\rm A}d_{\rm B} a\Omega \sin\phi/(24\ell^2)$,  
when the two spring constants are infinitely large $K_{\rm A}, K_{\rm B} \rightarrow \infty$ 
and $\lambda=1$. 
The third term in Eq.~(\ref{generalvelocity}) vanishes even if $d_{\rm A} \neq d_{\rm B}$ 
because $\hat \Omega \rightarrow 0$ holds in this limit. 
In the modified three-sphere swimmer model considered by Montino and DeSimone, one of the 
two arms was replaced by a passive elastic spring~\cite{Montino15}.
Their model can be obtained from the present model simply by setting one of the spring 
constants to be infinitely large, say $K_{\rm A} \rightarrow \infty$, and by regarding the 
natural length of the other spring as a constant, say $\ell_{\rm B}=\ell$ (or $d_{\rm B}=0$).
The continuous changes of the natural lengths introduced in Eqs.~(\ref{ellA}) and (\ref{ellB})  
are straightforward generalization of cyclically switched discrete states considered in the 
previous studies~\cite{GolestanianAjdari08,Golestanian10,Sakaue10,Huang12}.
We finally note that a similar model to the present one was considered in Ref.~\cite{Dunkel09},
although they focused only in the small-frequency region and did not discuss the entire 
frequency dependence.   
Using coupled Langevin equations, they mainly investigated  the interplay between self-driven 
motion and diffusive behavior~\cite{Dunkel09}, which is also an important aspect of microswimmers.

To summarize, we have discusses the locomotion of a generalized three-sphere microswimmer 
in which the spheres are connected by two elastic springs and the natural length of each spring 
is assumed to undergo a prescribed cyclic change.
As shown in Eqs.~(\ref{symvelocity}) and (\ref{generalvelocity}), we have analytically obtained 
the average swimming velocity $\overline V$ as a function of the frequency $\Omega$ of the 
cyclic change in the natural length.
In the low-frequency region, the swimming velocity increases with the frequency and reduces 
to the original three-sphere model by Najafi and Golestanian~\cite{Golestanian04,Golestanian08}.
In the high-frequency region, conversely,, the velocity is a decreasing function.
This property reflects the intrinsic spring relaxation dynamics of an elastic swimmer in a viscous fluid.

S.K.\ and R.O.\ acknowledge support from a Grant-in-Aid for Scientific Research on
Innovative Areas ``\textit{Fluctuation and Structure}" (Grant No.\ 25103010) from the Ministry
of Education, Culture, Sports, Science, and Technology of Japan and from 
a Grant-in-Aid for Scientific Research (C) (Grant No.\ 15K05250)
from the Japan Society for the Promotion of Science (JSPS).


\end{document}